\documentclass{article}
\usepackage{amssymb}
\usepackage{amsmath}

\input{tcilatex}
\begin{document}

\title{\textbf{Nonstandard interactions and interference effect in low
energy }$\nu _{e}e-$\textbf{scattering process}.}
\author{Amir Nawaz Khan\thanks{%
amir\_nawaz@comsats.edu.pk }, Farida Tahir\thanks{%
farida\_tahir@comsats.edu.pk } \\
\textit{Physics Department COMSATS\ Institute of\ InformationTechnology,} \\
\textit{Park Road Islamabad Pakistan.}}
\maketitle

\begin{abstract}
We study interference effect in$\ $elastic $\nu _{e}e-$scattering process in
presence of nonstandard neutrino interactions (NSI).The strength of
interference predicted by standard model (SM) is $-1.09,$ while that
measured in LAMPF and LSND experiment is $-1.01\pm 0.18$,\ which are in good
agreement with each other. We use interference effect (1) to invistigate how
NSIs could affect the total size of interference, (2) how interference can
be used to constrain NSIs and (3) how the allowed region for new physics can
be reduced from four to a single, but more symmetric allowed region.

\end{abstract}

\section{Introduction}

It is known that neutrinos are massless within the standard model, contrary
to this, plethora of neutrino oscillation experiments confirm that neutrinos
change their flavor while propagating from source to detector, thus
providing evidence that neutrios are massive\ \cite{sk, SNO, kamland, atm
new exp, k2k}\emph{. }General trend to accommodate the masses of neutrino
through effective four fermions operator is extensively discussed in
literature \cite{david, jb}. The effective operators approach provides a
plausible way to study the effects of new interactions in low energy
electroweak precision measurements. These new interactions are predicted by
various models such as R-parity violating supersymmetric model \cite{ftahir}%
, heavy neutral vector bosons $(Z^{^{\prime }})\ $model and the lepto-quarks
model \cite{david}.

The electroweak precision measurements have verified the universality and
flavor conserving processes of the SM \cite{elepre, andre}. But if neutrinos
are given masses then it is strongly suggested that it may have some new
interactions at some high energy scale $(\gtrsim 1TeV)\ $\cite{lwolf}. Such
interactions are called nonstandard neutrino interactions (NSIs). The NSIs
include nonuniversal flavor conserving as well as flavor changing currents
(also called flavor diagonal and flavor non-diagonal, respectively),
contrary to the standard model where both charge current (CC) and neutral
current (NC) are universal and flavor conserving. To explore new physics,
the study of nonuniversality and flavor violation couplings is of crucial
importance. Currently, constraining NSI coupling parameters, both in model
dependent and model idependent way, are extensively studied in electroweak
precision measurements \cite{elepre, andre}.

We study NSIs arising in neutrino-fermion scattering processes and focus on
the elastic $\nu _{e}e-$scattering. The major interest in this process is
due to the fact that it is one of the few processes for which SM predicts a
large distructive interference between CC and NC, thus provides the reason
for lowering the $\nu _{e}e$ cross section by 40\% compared to that in the
absence of interference. The presence of interference effect has been
discovered by CCFR neutrino experiment at Fermilab \cite{SRMISHRA} and
successively improved the results by LAMPF and short baseline terrestrial
LSND experiments \ \ \cite{allen1, allen2, allen3, lsnd}, and found good
agreements with the theory. The discovery of interference effect has
provided a crucial test for varifying the gauge structure of SM \cite%
{bkayser}. The inteference effect is deeply concerned with the fact that it
can occur if the incoming and out going neutrinos are the same in a
scattering process. For instance, in $\nu _{e}e-$scattering the interference
could occur only if the incoming and out going neutrinos are $\nu _{e}\ $%
\cite{allen1, allen2, allen3, lsnd}. We exploit this logic to study NSIs
using interference effect. This effect can be used both for constraining
NSIs in universal flavor conserving scatering and for knowing the guage
structure of any new physics. Based on the fact of flavor conservation
(incoming and out going neutrino should be the same) we can not ignore the
interference between CC, NC as estiblished in SM, and in addition, the
interference of NSIs with CC, NC. These facts impose us to reexamine the
strength of interference in the presence of NSI. If there exists any NSIs at
low energy, it must interfere with the CC and NC of the SM. More accurate
strength of total interference can be obtain using more and more accurate
measurements of $\nu _{e}e-$scatterings \cite{allen1, allen2, allen3, lsnd}.

\textit{An important aspect of the interferecnce parameter is that its
strength is energy independent. Whatever energy is used for scattering, the
total strength of interference is the same. Using this reasoning, we can use
interference effect as a probe to investigate any NSIs, if exist. The
impacts of interference between SM interactions and NSIs\ has been discussed
in ref. \cite{phuber}. It is shown in this ref.\cite{phuber} that how a
small residual NSIs could interfere with SM interacations and leads to a
drastic loss in sensitivity in }$\theta _{13}.$

\textit{In this paper, we investigate NSIs using interference as parameter
in the low energy }$\nu _{e}e-$scattering process.\textit{\ The same
analysis can be performed for the other scattering processes like }$\nu
_{\mu }e\ $and $\nu _{\tau }e\ $\footnote{%
Work in progress}\ . W\textit{e demonstrate how interference parameter can
be used to constrain NSIs. We obtain new bounds on NSIs using interference
parameter following the approach of keeping one operator at a time. The
lower bounds on }$\epsilon _{ee}^{eL}$\textit{\ and }$\epsilon _{ee}^{eR}\ $%
\textit{obtained are more} \textit{stringent in our case where as upper
bounds relax the allowed region. These bounds are complimentary to one or
another obtained using various different methods in ref \cite{david, jb}. On
the other hand, we obtain a single allowed region instead of four. Four
allowed regions are obtained when neutinos and anti-neutrino data is
similtaneously used. This analysis has recently been done by J.Barranco et.
al. (see ref. \cite{jb}). Our approach helps to take into account the two
parameters instead of single parameter at a time which is commonly followed
in the literature (see ref. \cite{david, jb}).}

\section{NSI Lagrangian}

The most general form of the effective four-fermion interaction Lagrangian
for low energy $(\nu _{\alpha }f\longrightarrow \nu _{\beta }f\ )$ process
in the presence of NSI is given by \cite{david},

\begin{eqnarray}
\tciLaplace ^{eff} &=&-2\sqrt{2}G_{f}[\bar{\nu}_{\alpha }\gamma _{\mu
}Ll_{\alpha }][\bar{f}\gamma ^{\mu }Pf]-2\sqrt{2}\underset{p,f,\alpha }{%
\dsum }g_{p}^{f}G_{f}\ [\bar{\nu}_{\alpha }\gamma _{\mu }L\nu _{\alpha }][%
\bar{f}\gamma ^{\mu }Pf\ ]  \notag \\
&&-\underset{\alpha ,\beta }{\dsum }\epsilon _{\alpha \beta }^{fp}2\sqrt{2}%
G_{f}[\bar{\nu}_{\alpha }\gamma _{\mu }L\nu _{\beta }][\bar{f}\gamma ^{\mu
}Pf]
\end{eqnarray}%
where $P=L,R=\frac{1}{2}\ (1\mp \gamma _{5})$ with $G_{F}$\ as the Fermi
constant and\ $f$ is any of the first generation fermion $(e,u,d)$, $%
g_{p}^{f}\ $are the standard neutral current coupling constants and $%
\epsilon _{\alpha \beta }^{fp}\ $are the nonstandard flavor diagonal\textbf{%
\ }$(\alpha =\beta )${\large \ }and flavor nondiagonal $(\alpha \neq \beta
)\ $effective coupling parameters.

For the specific process of $\nu _{e}e-$scattering,$\ $the total effective
Lagrangian becomes,

\begin{eqnarray}
\tciLaplace ^{eff} &=&-2\sqrt{2}G_{f}[\bar{\nu}_{e}\gamma _{\mu }L\nu _{e}][%
\bar{e}\ \gamma ^{\mu }Pe]-2\sqrt{2}g_{p}^{f}G_{f}\ [\bar{\nu}_{e}\gamma
_{\mu }L\nu _{e}][\bar{e}\ \gamma ^{\mu }Pe]  \notag \\
&&-\epsilon _{ee}^{ep}2\sqrt{2}G_{f}[\bar{\nu}_{e}\gamma _{\mu }L\nu _{e}][%
\bar{e}\gamma ^{\mu }Pe]
\end{eqnarray}%
\newline
Notice that first and third terms have been obatined in this form after
Fierz rearrangement. For the detail on effecttive lagrangian formalism of
NSIs see ref. \cite{david, jb}.

\section{Interference effect in SM and measurements}

Using the standard model part of\ lagrangian in $(2),$ the total cross
section can be calculated as

\begin{equation}
\sigma ^{\nu _{e}}=\sigma ^{o}[(g_{L}+2)^{2}+\frac{g_{R}^{2}}{3}]
\end{equation}%
where $\sigma ^{o}=\frac{G_{f}^{2}m_{e}E_{\nu _{e}}}{2\pi }=(4.31\times
10^{-45})\frac{cm^{2}}{MeV}\times E_{\nu e},\ \ \ g_{_{L}}=-1+2\sin \theta
_{w}^{2}\ $and\ $g_{_{R}}=2\sin ^{2}\theta _{w}.$

To make the interference term more explicit, we rewrite $(3)$ in the form,

\begin{equation}
\sigma ^{\nu e}=\sigma ^{CC}+\sigma ^{NC}+\sigma ^{I}
\end{equation}%
\ \ where$\ \sigma ^{CC}=4\sigma ^{o},\ \ \sigma ^{NC}=\sigma
^{o}(g_{L}{}^{2}+\frac{g_{R}^{2}}{3}),\ \ \sigma ^{I}=4\sigma
^{o}g_{L}=2\sigma ^{o}I^{SM}\ ,$

where%
\begin{eqnarray}
I^{SM} &=&2g_{L}  \notag \\
&=&2(-1+2\sin \theta _{w}^{2})
\end{eqnarray}%
Assuming $\sin ^{2}\theta _{w}=0.23,\ \sigma ^{NC}\ $and $\sigma ^{I}$\ can
calculated within the SM as $\sigma ^{NC}=0.36\sigma ^{o},\ \sigma ^{I}\
=2(-1.1)\sigma ^{o}\ $where $I^{SM}=2g_{L}=-1.1\ $\cite{lsnd}$.$

Including the radiative corrections \cite{lsnd, redcor}, we obtain $\ \sigma
^{NC}=0.37\sigma ^{o}\ $and\ $\sigma ^{I}=2(-1.09)\sigma ^{o}$ with $%
I^{SM}=-1.09.$ The total cross section becomes,

\begin{equation*}
\sigma ^{\nu _{e}e}=4\sigma ^{0}+0.37\sigma ^{0}+2(-1.09)\sigma ^{0}
\end{equation*}

From the third term, it is clear that the standard model predicts
destructive interference between $CC$ and $NC$ having absolute value of $%
1.09 $.

\textit{From eq. (5), we can see that interference between CC and NC is a
function of the weak mixing angle }$\theta _{W}.$\textit{The strength of
interferece in SM is -1.09 corresponds to 0.5 radian (for }$\sin ^{2}\theta
_{w}=0.23$\textit{) of \ }$\theta _{W}$\textit{. We can see from figure (1),
that the maximum size of destructive interference corresponds to -2 \ and it
vanishes at 0.8 radian and beyond this we have constructive interference.
Although, at the SM energy scale the physical size of interference is -1.09
and the remaining is the unphysical region, but these information which is
deduced from the nature of interferecne can be used to test the gauge
structure of any interaction beyond the SM.}

Now for experimental measurment of the size of interference we have from eq.
(4)$,$%
\begin{eqnarray}
I &=&\frac{\sigma ^{\nu e}-(\sigma ^{CC}+\sigma ^{NC})}{2\sigma ^{0}}  \notag
\\
I &=&\frac{\sigma _{\exp }}{2\sigma ^{0}}-2.185
\end{eqnarray}%
where $\sigma ^{\nu e}\equiv \sigma _{\exp }\ $and $\sigma ^{CC}=4\sigma
^{o},\ \sigma ^{NC}=0.37\sigma ^{o}$ were used to obtain the eq. (6).

Using $\sigma _{\exp }=[10.1\pm 1.1(stat.)\pm 1.0(syst.)\times E_{\nu
e}(MeV)\times 10^{-45}cm^{2}]$ in eq.(4) from the LSND\ experiment \cite%
{lsnd} and solving for $I,$ we get $I\ ^{LSND}=-1.01\pm 0.18.$ Comparing $%
I^{SM}\ $and $I\ ^{LSND},$ one can see a discrepency of $0.08\ $which is $%
8\%\ $with respect to the best value of LSND\ experiment. The destructive
interference(-ev sign) is in agreement with both, the theory and experiment.

\textit{Note that for the experimental measurement of interference, the
inputs for CC and NC cross sections were taken from separate experiments for
purely leptonic processes. For CC, muon decay measurement was taken and for
NC, }$\nu _{\mu }e-$scattering measurement were used \cite{allen1, allen2,
allen3, lsnd}. \textit{\ }

Inspite of this agreement between theory and experiment for the stregnth of
interference we can not ignore impact of NSIs (if there exist any due to
massive neutrinos) on the total size and sign of interference.This is
because of the fact that in the total interference term some currents may
interact constructively and some distructively which cancel each others
effect and thus the over all size remain the same or may change by a small
amount, which in turn make the total cross section as unchanged.

In the following section, we follow the same approach as adopted in \cite%
{allen1, allen2, allen3, lsnd},(1) to invistigate how NSIs could affect the
total size of interference, (2) how interference can be used to constrain
NSIs and how the allowed region for new physics can be reduced from four to
a single, but more symmetric allowed region.

\section{Interference effect and NSIs}

In the presence of NSIs, total cross section calculated as \cite{jb}

\begin{equation}
\sigma ^{\nu _{e}e}=\sigma ^{o}[\tilde{g}_{L}{}^{2}+\frac{\tilde{g}_{R}^{\ 2}%
}{3}]
\end{equation}%
where%
\begin{equation*}
\tilde{g}_{_{L}}^{^{\ }}=2+g_{L}+\epsilon _{ee}^{eL},\ \tilde{g}%
_{_{R}}=g_{R}+\epsilon _{ee}^{eR},\ g_{_{L}}=-0.54,\ g_{_{R}}=0.46
\end{equation*}
Rewritting $(7)\ $as,

\begin{equation}
\sigma ^{\nu _{e}e}=\sigma ^{CC}+\sigma ^{NC}+\sigma ^{NSI}+\sigma ^{I}
\end{equation}
where%
\begin{eqnarray*}
\sigma ^{CC} &=&4\sigma ^{o},\sigma ^{NC}=(g_{L}^{2}+\frac{1}{3}%
g_{R}^{2})\sigma ^{o}=0.37\sigma ^{o},\sigma ^{NSI}=\{(\epsilon
_{ee}^{eL})^{2}+\frac{1}{3}(\epsilon _{ee}^{eR})^{2}\}\sigma ^{o} \\
\ \sigma ^{I}\ \ &=&2\{(2g_{L}+g_{L}(\epsilon _{ee}^{eL})+2(\epsilon
_{ee}^{eL})+\frac{1}{3}g_{R}(\epsilon _{ee}^{eR})\}\sigma
^{o}=2I^{total}\sigma ^{o}\ 
\end{eqnarray*}%
where 
\begin{equation}
I^{total}=\{(2g_{L}+g_{L}(\epsilon _{ee}^{eL})+2(\epsilon _{ee}^{eL})+\frac{1%
}{3}g_{R}(\epsilon _{ee}^{eR})\}
\end{equation}

\textit{In eq. (8)}$,$\textit{\ the first term is the SM interference term,
the second and fourth are interference terms between NC and NSIs and third
term is the interference between the CC and NSIs. One important point which
is noticeable is that from fourth term where the interference between right
handed coupling constant of SM\ }$(g_{R})\ $\textit{and right \ handed
coupling parameter of NSI (}$\epsilon _{ee}^{eR}$) occurs,\textit{\ while
contrary to this, the interference in the SM is only between the CC and NC}$%
.\ $There is no interference due to the right handed part of NC$\ $in SM.

Substituting $g_{L}=-0.54\ $and $g_{R}=0.46,\ $the total interference ($%
I^{total}$) can be written as,%
\begin{equation}
I^{total}=-1.09+1.46(\epsilon _{ee}^{eL})+0.15(\epsilon _{ee}^{eR})
\end{equation}

T\textit{he first term which is the SM interference term is obviously
destructive while the sign of the second and third terms, which are NSI
terms, depends on the signs of }$\epsilon _{ee}^{eL}$\textit{\ and }$%
\epsilon _{ee}^{eR}$. If these are negative, the interference will be
destructive and if the signs are positive there will be constructive
interference.

Using single parameter approach, (considering one operater at a time) we get
bounds from the interference term for the measured value of LSND ($I\
^{LSND}=-1.01\pm 0.18$)%
\begin{eqnarray}
-0.07 &<&\epsilon _{ee}^{eL}<0.17\   \notag \\
-0.35 &<&\epsilon _{ee}^{eR}<0.81
\end{eqnarray}

If we assume that the discrepancy between theory and experimental size of
interference, which is $0.08\ $comes from NSIs then using eq. $(8),\ $and
single parameter a time we find $\epsilon _{ee}^{eL}=0.05\ $ $\epsilon
_{ee}^{eR}=0.52.$\ Both from these absolute values of $\epsilon _{ee}^{eL}\ $%
and $\epsilon _{ee}^{eR},\ $and bounds in eq. (10) it clear that the right
hand NSIs parameters have largers values. The upper bound on $\epsilon
_{ee}^{eL}$ are quite compitable with that obtained before in \cite{david,
jb}. The interference\ ($I$) as a function of weak mixing angle $\theta
_{W},\ $NSI parameters ($\epsilon _{ee}^{eL}\ $and$\ \epsilon _{ee}^{eR})$
is shown in fig (2). In case of NSI parameters $(\epsilon _{ee}^{eL}\ $and$\
\epsilon _{ee}^{eR}),\ $one parameter at a time was considered, while the
other parameter is kept zero.

\ If NSIs is taken into consideration and if they interfere with SM\
currents, then the total interference ($I$) in any experiment will be
modified from eq. (6) to the form,

\begin{eqnarray}
I &=&\frac{\sigma ^{\nu e}-(\sigma ^{CC}+\sigma ^{NC}+\sigma ^{NSI})}{%
2\sigma ^{0}}  \notag \\
\ I &=&\frac{\sigma _{\exp }}{2\sigma ^{0}}-2.185-\{(\epsilon
_{ee}^{eL})^{2}+\frac{1}{3}(\epsilon _{ee}^{eR})^{2}\}
\end{eqnarray}

where\ $\sigma ^{NSI}=\{(\epsilon _{ee}^{eL})^{2}+\frac{1}{3}(\epsilon
_{ee}^{eR})^{2}\}\sigma ^{o}$

For $I=I^{SM}=-1.09\ $and $\sigma _{\exp }=[10.1\pm 1.1(stat.)\pm
1.0(syst.)\times E_{\nu e}(MeV)\times 10^{-45}cm^{2}]$ from LSND\ experiment 
\cite{lsnd} and $\sigma ^{o}=(4.31\times 10^{-45})\frac{cm^{2}}{MeV}\times
E_{\nu e}$, eq. (11) becomes,

\begin{equation}
(\epsilon _{ee}^{eL})^{2}+\frac{1}{3}(\epsilon _{ee}^{eR})^{2}=(0.26\uparrow
,\ -0.1\downarrow )
\end{equation}

This has been plotted in figure below.

$\ $

The most important feature of using interference parameter for contraining
NSIs is to look for the overlaped region between the total cross section eq.
(8) and the interference term eq. (10). We get a single overlaped region,
which is more symmetric with respect to the lower and upper bounds. Already,
analysis has been done to get overlaped regions using $\nu _{e}e~$and $%
\overline{\nu _{e}}e$ -scattering data \cite{jb}. In that case four allowed
regions were obtained. Our analysis reduces the four allowed regions to a
single more symmetric allowed region as shown in figure $(4)$. The allowed
region is bounded by the limits:%
\begin{eqnarray*}
-0.25 &<&\epsilon _{ee}^{eL}<0.25 \\
-1.65 &<&\epsilon _{ee}^{eR}<1.65
\end{eqnarray*}

\section{Conclusions}

The interference effect between CC and NC \ in $\nu _{e}e-$scattering
process has been observed in the standard model. The size of this
interference in the SM is $-1.09$, whereas that measured in LSND\ experiment
is $-1.01\pm 0.18.\ $The theory vs experiment discrepancy is $8\%$.$\ $Here
we have reanalysed the interference effect to use it as probe for NSIs. We
used the interference effect to invistigate how NSIs could affect the total
size of interference, how interference can be used to constrain NSIs and how
the allowed region for new physics can be reduced from four to one single,
but more symmetric allowed region.

\section{Acknowledgement{}}

Farida Tahir thanks Prof. Douglas McKay for stimulating her interest in this
Physics. She is also thankful to Prof. John Ellis for providing her with an
opportunity to work at CERN. She is also grateful to S. Pascoli and E. F.
Martinez for their valuable discussions. This work was supported by Higher
Education Commission (HEC) of Pakistan under grant$-\#\ 20-577$ and the
indigenous PhD $5000$ Fellowship program phase$-IV$.

\end{document}